\documentclass[11pt]{article}
\usepackage{amsmath}
\usepackage{graphicx}
\usepackage{dcolumn}
\usepackage{bm}
\usepackage{wrapfig}
\usepackage{epsfig}
\usepackage{breqn}
\usepackage{amssymb}
\usepackage{subfigure}
\usepackage[usenames,dvipsnames]{pstricks}
\usepackage{epsfig}
\usepackage{pst-grad} 
\usepackage{pst-plot} 
 \usepackage[usenames,dvipsnames]{pstricks}
 \usepackage{epsfig}
 \usepackage{pst-grad} 
 \usepackage{pst-plot} 
 \usepackage{float}

\begin{document}

\title{On Singularities in Cosmic Inflation}
\author{Ikjyot Singh Kohli \\isk@mathstat.yorku.ca \\York University - Department of Mathematics and Statistics}
\date{April 25, 2016}                                           

\maketitle

\begin{abstract}
In this paper, we examine a flat FLRW spacetime with a scalar field potential and show by applying Osgood's criterion to the Einstein field equations that all such models, irrespective of the particular choice of potential develop finite-time singularities. That is, we show that solutions to the field equations become singular in finite time, which can have important implications for the role of inflation in cosmological models. We further point out that a possible reason for this behaviour is that the solutions to the field equations in such inflationary scenarios do not obey global existence and uniqueness properties, which is a typical characteristic of solutions that diverge in finite time. 
\end{abstract}

\section{Introduction}
The purpose of this paper is to demonstrate the existence of future singularities that occur in finite time in inflationary models with an arbitrary scalar field potential within the context of a spatially flat Friedmann-Lema\^{i}tre-Robertson-Walker (FLRW) spacetime. The concept of the existence of a singularity in the context of general relativity has been studied quite extensively. With respect to the cosmological case, there is the fundamental singularity theorem \cite{elliscosmo} \cite{tolmanward} \cite{raych1955} which describes irrotational geodesic singularities, and states that if $\Lambda \leq 0$, $\mu + 3p \geq 0$, and $\mu + p > 0$ in a fluid flow where $\mu$ is the energy density of the fluid, $p$ is the pressure of the fluid, in addition to having $\dot{u} = 0$, $\omega = 0$, and $H_{0} = 0$ at some time $s_{0}$, then a spacetime singularity, where either the expansion scalar goes to zero or the shear scalar diverges occurs at a finite proper time $\tau_{0} \leq 1/H_{0}$ before $s_{0}$.

Further elaborations are built upon this singularity theorem. It is of interest to note that in \cite{elliscosmo} five possible routes to avoid the conclusions of this singularity theorem are discussed in detail. They are a positive cosmological constant, acceleration, vorticity, an energy condition violation, or alternative gravitational equations. We refer the interested reader to \cite{elliscosmo} for further discussions regarding these issues. 

Following \cite{elliscosmo}, we also note that singularities occur in cosmology not only in the context of FLRW models, but also for realistic anisotropic and inhomogeneous models of the universe in which the strong energy condition $\mu + 3p > 0$ is satisfied. Related to this, Penrose \cite{penrose1965} did pioneering work on black hole singularities producing a theorem that allowed one to predict the existence of singularities in realistic gravitational collapse cases. Hawking \cite{penrosehawking1970} extended these results to the cosmological context leading to the famous Hawking-Penrose singularity theorem which implied that space-time singularities are to be expected if either the universe is spatially closed or there is an `object' undergoing relativistic gravitational collapse (existence of a trapped surface) or there is a point $p$ whose past null cone encounters sufficient matter that the divergence of the null rays through $p$ changes sign somewhere to the past of $p$ (i.e. there is a minimum apparent solid angle, as viewed from $p$ for small objects of given size). The theorem applies if the following four physical assumptions are made: (i) Einstein's equations hold (with zero or negative cosmological constant), (ii) the energy density is nowhere less than minus each principal pressure nor less than minus the sum of the three principal pressures (the `energy condition'), (iii) there are no closed timelike curves, (iv) every timelike or null geodesic enters a region where the curvature is not specially aligned with the geodesic. (This last condition would hold in any sufficiently general physically realistic model.) 

Further, Bekenstein \cite{bekenstein1974} studied exact solutions of Einstein-conformal scalar equations and presented a class of FLRW models which contained both incoherent radiation and a homogeneous conformal scalar field which bounce and never pass through a singular state, thereby circumventing the singularity theorems by violating the energy condition. Parker and Fulling \cite{parkerfulling1973} considered a classical gravitational field minimally coupled to a quantized neutral scalar field possessing mass. They concluded that quantum effects can sometimes lead to the avoidance of the cosmological singularity, at least on the time scale of one Friedmann expansion. Collins and Ellis \cite{collinsellis1979} examined in detail the singularities that occur in Bianchi cosmologies.
Barrow and Matzner \cite{barrowmatz1977} showed that the singularity corresponding to homogeneous and isotropic universes observationally equivalent to ours must be of simultaneous Robertson-Walker type containing only small curvature fluctuations. In an interesting application of Robertson-Walker singularities, Barrow \cite{barrow1977} showed that the large-scale velocity and vorticity fields required for the vortex and spinning-core theories of galaxy formation can be generated primordially in a natural way from a FLRW singularity.

This paper in particular is concerned with the existence of singularities in inflationary models. Some work has been done in this regard. In particular, Kohli and Haslam \cite{kohlihaslam} proved the existence of finite-time singularities in the case of stochastic eternal inflation for an arbitrary inflaton field potential. Nojiri, Odintsov, and Oikonomou \cite{nojiri1} analyzed in detail the concept of singular inflation with scalar-tensor and modified gravity. Nojiri, Odintsov, Oikonomou and Saridakis \cite{nojiri2}  studied the existence of finite-time singularities in the inflationary and late-time regimes of cosmological evolution. Nojiri, Odintsov, and Tsujikawa \cite{nojiri3} investigated in detail the properties of future singularities in a universe dominated by dark energy including a phantom-type fluid. Recently, Odintsov and Oikonomou \cite{nojiri4} examined cases in which singular evolution can be consistently incorporated in deformations of the $R^2$ inflationary potential. Barrow and Graham \cite{barrowgraham} proved that a spatially homogeneous and isotropic universe containing a scalar field with potential $V(\phi) = A \phi^{n}$ with $0<n<1$ and $A>0$ always develops a finite-time singularity at which the Hubble rate and its first derivative are finite, but its second derivative diverges. We are not aware of any other work in the literature that describes the existence of finite-time singularities in FLRW inflationary cosmologies. Note that throughout we used geometrized units where $8\pi G = c = 1$.

\section{Description of the Model}
We consider a spatially homogeneous and isotropic universe described by the FLRW metric \cite{elliscosmo} as
\begin{equation}
\label{FLRW1}
ds^2 = -dt^2 + a^2(t) \left[dr^2 + f^2(r) \left(d\theta^2 + \sin^2 \theta d \phi^2\right)\right], \quad u^\mu = \delta^{\mu}_{0},
\end{equation}
where 
\begin{equation}
f(r) = (\sinh r, r, sin r) \mbox{ for } K = (-1, 0, +1),
\end{equation}
where $K$ denotes the sign of the curvature of the particular FLRW model under consideration. Namely, $K = -1$ refers to hyperbolic FLRW models, $K = 0$ refers to flat FLRW models, and $K=+1$ refers to positively curved FLRW models.

We assume that energy-momentum tensor is that of a scalar field and has the form \cite{elliscosmo}
\begin{equation}
\label{emtensor}
T_{\phi}^{ab} = \nabla^{a} \phi \nabla^{b} \phi - \left[\frac{1}{2} \nabla_{c} \phi \nabla^{c} \phi + V(\phi)\right]g^{ab}.
\end{equation}

The Einstein field equations that also describe the dynamics of the model are given by the Raychaudhuri and Friedmann equations \cite{elliscosmo},
\begin{eqnarray}
\label{raych}
\dot{H} &=& -H^2 + \frac{1}{3} \left[V(\phi) - \dot{\phi}^2\right],  \\
\label{fried1}
3H^2 &=& V(\phi) + \frac{\dot{\phi}^2}{2} + \frac{1}{2} ^{(3)}R,
\end{eqnarray}
where $H$ is the Hubble parameter, and $^{(3)}R$ is the three-dimensional Ricci scalar, which is constant for the FLRW models.

The contracted Bianchi identities give an evolution equation for the scalar field, $\phi$, which in this case is precisely the Klein-Gordon equation \cite{elliscosmo},
\begin{equation}
\label{kgordon}
\ddot{\phi} + 3H\dot{\phi} + V'(\phi) = 0.
\end{equation}

Together Eqs. \eqref{raych}, \eqref{fried1}, and \eqref{kgordon} fully describe the dynamics of the cosmological model described by Eqs. \eqref{FLRW1} and \eqref{emtensor}.

We further define expansion-normalized variables denoted by $X,Y$ as
\begin{equation}
\label{eq:Xdef}
X = \sqrt{\frac{V}{3}}\frac{1}{H}, \quad Y = \frac{\dot{\phi}}{\sqrt{6} H}.
\end{equation}
As is done in \cite{ellis}, we additionally introduce a dimensionless time variable $\tau$, such that
\begin{equation}
\label{eq:dimtime}
\frac{dt}{d\tau} = \frac{1}{H}.
\end{equation}

Substituting Eqs. \eqref{eq:Xdef} and \eqref{eq:dimtime} into Eqs. \eqref{raych}-\eqref{kgordon}, one obtains the following dynamical system
\begin{eqnarray}
\label{eq:Xev}
X' &=& X \left(1 + q + Y \lambda\right), \\
\label{eq:Yev}
Y' &=& - 3 Y - \frac{1}{3} \left(1-Y^2\right) + Y\left(1+q\right),
\end{eqnarray}
subject to the constraint
\begin{equation}
\label{eq:constr2}
X^2 + Y^2 = 1.
\end{equation}
Note that $q$ is the deceleration parameter found through Raychaudhuri's equation and is in this case
\begin{equation}
\label{eq:qdef}
q = 2Y^2 - X^2.
\end{equation}

Following \cite{wagstaff}, we also have defined a model-dependent dimensionless parameter,
\begin{equation}
\label{eq:lambda}
\lambda = \sqrt{\frac{3}{2}} \frac{V'}{V}.
\end{equation}
Note that in the definition of $\lambda$ in Eq. \eqref{eq:lambda}, we have set the Planck mass to unity as per our choice of units.  This will allow us to consider the dynamics corresponding to an arbitrary inflaton potential making our results as general as possible.

Making use of Eqs. \eqref{eq:constr2} and \eqref{eq:qdef} in Eqs. \eqref{eq:Xev} and \eqref{eq:Yev}, we see that the equations decouple,
\begin{eqnarray}
\label{eq:Xev1}
X' &=& X \left(3 - 3X^2 + \sqrt{1 - X^2}\lambda\right), \\
\label{eq:Yev1}
Y' &=& \left(Y^2-1\right)\left(3 Y  + \lambda\right).
\end{eqnarray}

Notice that from Eq. \eqref{emtensor}, that the energy-momentum tensor is that of a perfect fluid with energy density
\begin{equation}
\label{eq:mudef}
\mu = \frac{1}{2}\dot{\phi}^2 + V(\phi),
\end{equation}
and pressure
\begin{equation}
\label{eq:pdef}
p = \frac{1}{2}\dot{\phi}^2 - V(\phi).
\end{equation}
Following \cite{elliscosmo}, we note that for a non-negative potential energy $V \geq 0$, one has the restriction that 
\begin{equation}
\label{eq:restr1}
\frac{p}{\mu} \leq 1.
\end{equation}
%

For what follows, we will exclusively concentrate on the dynamics represented by Eq. \eqref{eq:Xev1}. That is, we will analyze the initial value problem
\begin{eqnarray}
X' &=& X \left(3 - 3X^2 + \sqrt{1 - X^2}\lambda\right), \nonumber \\
X(0) &=& \xi.
\end{eqnarray}

\section{Osgood's Criterion and Finite-Time Singularities}
We will demonstrate the existence of finite-time singularities by making use of Osgood's criterion applied to Eq. \eqref{eq:Xev1}. Osgood's criterion due to W.F. Osgood (1898) \cite{osgood2} which is a very fundamental result describing conditions for the uniqueness of solutions to ordinary differential equations. As noted in \cite{leonja}, Osgood established that the solution $y(t)$ of
\begin{eqnarray}
y'(t) &=& b\left(y(t)\right), \quad t > 0, \\
y(0) &=& \xi, \nonumber
\end{eqnarray}
diverges within a \emph{finite} time, that is, $y(t) \to \pm \infty$ if
\begin{equation}
\label{eq:integral1}
\int_{\xi}^{\infty} \frac{ds}{b(s)} < \infty.
\end{equation}
We further note that the time it takes for the solution to become singular is given by
\begin{equation}
\label{eq:Tdef}
T = \int_{\xi}^{\infty} \frac{ds}{b(s)},
\end{equation}
such that
\begin{equation}
T = \sup \left\{\tau > 0: T < \infty \right\}.
\end{equation}

Motivated by Osgood's criterion, recalling Eq. \eqref{eq:Xev1}, we denote
\begin{equation}
\label{eq:F}
F(X) = X \left(3 - 3X^2 + \sqrt{1 - X^2}\lambda\right).
\end{equation}

We now compute the integral in \eqref{eq:integral1} using Eq. \eqref{eq:F}, and obtain
\begin{eqnarray}
\int_{\xi}^{\infty} \frac{dS}{F(S)} &=& \int_{\xi}^{\infty} \frac{dS}{S \left(3 - 3S^2 + \sqrt{1 - S^2}\lambda\right)} \nonumber \\
\label{eq:integral2}
&=& \frac{3 \log 3}{\lambda^2 - 9} - \frac{\log \left|1-\sqrt{1-\xi^2}\right|}{2(3+\lambda)} - \frac{\log \left|1 + \sqrt{1-\xi^2}\right|}{2(3-\lambda)} + \frac{3 \log \left|3\sqrt{1-\xi^2} + \lambda\right|}{9-\lambda^2}. \nonumber \\
\end{eqnarray}
where $X(0) = \xi$.

The integral as computed in Eq. \eqref{eq:integral2} diverges if $\lambda = \pm 3$. That is, per Osgood's criterion and Eq. \eqref{eq:lambda}, there is no finite-time singularity if and only if
\begin{equation}
\label{eq:choice1}
\sqrt{\frac{3}{2}} \frac{V'(\phi)}{V(\phi)} = \pm 3.
\end{equation}

Except for the specific choice of the inflaton potential as described in Eq. \eqref{eq:choice1}, the types of inflationary models as described in this paper always admit a finite-time singularity, where $X(\tau) \to \infty$, for $\tau < \infty$.

\section{Discussion}
We have shown thus far that inflationary models with arbitrary potential $V(\phi)$ in a flat FLRW background become singular in a finite time. Exploring this fact in more detail, recalling Eq. \eqref{eq:F}, we note that
\begin{equation}
\label{eq:Fprime}
F'(X) = X^2 \left(-\frac{\lambda }{\sqrt{1-X^2}}-9\right)+\lambda  \sqrt{1-X^2}+3.
\end{equation}
One therefore sees that $F'(X)$ is only continuous for any subinterval of $-1 < X < 1$, albeit, it is also the domain for $F'(X)$. This implies that $F'(X)$ is also bounded over any closed subinterval of $-1<X<1$, that is, $F(X)$ is locally Lipschitz continuous over this domain. This implies by the existence and uniqueness theorem of ordinary differential equations \cite{boyce}, that over any subinterval of $-1<X<1$, solutions to Eq. \eqref{eq:Xev1} will exist and be unique. 

However, the computation of the integral in Osgood's criterion in Eq. \eqref{eq:integral2} also shows that while $X(\tau) \to \infty$ in finite time, it will first exit the domain of definition in finite time, that is, $X(\tau) \to 1$ in finite time. This is the point at which the singular solution begins to develop. Since, at this point, one can see that Eq. \eqref{eq:Fprime} diverges, and so, all one has is the continuity of $F(X)$ at $X=1$, which only guarantees existence of solutions, but not uniqueness. For the Einstein field equations to be well-posed, one needs both existence and uniqueness of solutions, which is clearly not the case for the inflationary models described in this paper. 

\section{Conclusions}
In this paper, we examined a flat FLRW spacetime with a scalar field potential and showed through Osgood's criterion that all such models, irrespective of the particular choice of potential develop finite-time singularities, which is due to the fact that solutions to the field equations rapidly diverge in a finite time. We further discussed that a possible reason for this behaviour is that the solutions to the field equations in inflationary models do not obey global existence and uniqueness, which is a typical characteristic of solutions that diverge in finite time. This, as it turns out, is a very important point. 

Even if solutions to a particular differential equation are unique over a finite interval, one can continuously patch together such intervals to form a maximal interval on which the solution is defined. That is,  let $\psi_{a}(\tau)$ be the unique solution of the differential equation in question, which satisfies $\psi_{a}(0) = a$, and let $(\tau_{min}, \tau_{max})$ denote the maximal interval on which $\psi_{a}(t)$ is defined. If $t_{max}$ is finite, then \cite{ellis}
\begin{equation}
\lim_{\tau \to \tau_{max}} \left\| \psi_{a}(\tau) \right\| = +\infty.
\end{equation}
That is, solutions can \emph{only} be extended indefinitely unless the solution diverges in a finite time. However, we showed in this paper, that this is not the case, as solutions indeed diverge in a finite time.

\section{Acknowledgements}
ISK would like to thank Jorge A. Le\'{o}n for interesting discussions regarding the applications of Osgood's criterion to ordinary differential equations. ISK would also like to thank Vasilis Oikonomou and Emmanuel Saridakis for providing additional insights and references regarding this work. 

\newpage
\bibliographystyle{ieeetr} 
\bibliography{sources}

\end{document}